\documentclass[preprint,showpacs,preprintnumbers,amsmath,amssymb,nofootinbib]{revtex4}

\usepackage{etex}
\usepackage{amssymb,amsthm,amscd,amsbsy,array}
\usepackage{bm}
\usepackage{soul} 
\usepackage{graphics,graphicx,xcolor}

\usepackage[colorlinks=true, pdfstartview=FitV, linkcolor=blue, citecolor=blue, urlcolor=blue]{hyperref} 

\newcommand{\red}{\textcolor{red}}

\newcommand{\blue}{\textcolor{blue}}

\newcommand{\gb}{\quad\colorbox{green}}

\newenvironment{redtext}{\color{red}}
{\ignorespacesafterend}
\newenvironment{bluetext}{\color{blue}}{\ignorespacesafterend}

\newenvironment{magentatext}{\color{magenta}}{\ignorespacesafterend}
\newenvironment{cyantext}{\color{cyan}}{\ignorespacesafterend}
\newenvironment{orangetext}{\color{orange}}
{\ignorespacesafterend}

\newcommand{\bmagenta}{\begin{magentatext}}
\newcommand{\emagenta}{\end{magentatext}}
\newcommand{\bcyan}{\begin{cyantext}}
\newcommand{\ecyan}{\end{cyantext}}
\newcommand{\bblue}{\begin{bluetext}}
\newcommand{\eblue}{\end{bluetext}}
\newcommand{\bred}{\begin{redtext}}
\newcommand{\ered}{\end{redtext}}
\newcommand{\borange}{\begin{orangetext}}
\newcommand{\eorange}{\end{orangetext}}

\numberwithin{equation}{section}

\let\ssection=\section
\renewcommand{\section}{\setcounter{equation}{0}\ssection}
\newcommand{\beq}{\begin{equation}}
\newcommand{\eeq}{\end{equation}}

\def\JI{{Junker-Inomata\;}}

\def\GW{{gravitational wave\,}}

\newcommand{\bA}{{\mathbf{A}}}

\newcommand{\cA}{{\mathcal{A}}}

\newcommand{\calS}{{\mathcal{S}}}
\newcommand{\calA}{{\mathcal{A}}}

\newcommand{\bx}{{\bm{x}}}
\newcommand{\bX}{{\bm{X}}}
\newcommand{\bxi}{{\bm{\xi}}}
\newcommand{\bzeta}{{\bm{\zeta}}}
\newcommand{\baromega}{\bar{\omega}}

\def\aand{{\quad\text{\and and}\quad}}
\def\where{{\quad\text{\small where}\quad}}
\def\with{{\quad\text{\small with}\quad}}

\newcommand{\bF}{{\bf F}}

\newcommand{\bgamma}{\boldsymbol{\gamma}}

\def\smallover\#1/\#2{\hbox{$\textstyle\frac{\#1}{\#2}$}} %

\def\JI{{Junker and Inomata\,}}

\def\parag{\hfil\break} 
\def\kikezd{\parag\underbar}

\def\bequ{\begin{enumerate}}
\def\eenu{\end{enumerate}}
\def\bitem{\begin{itemize}}
\def\eitem{\end{itemize}}

\def\beq{\begin{equation}}
\def\eeq{\end{equation}}
\def\beqa{\begin{eqnarray}}
\def\eeqa{\end{eqnarray}}

\def\barray{\left(\begin{array}}
\def\earray{\end{array}\right)}
\def\barraynb{\begin{array}}
\def\earraynb{\end{array}}

\def\GW{{gravitational wave\;}}
\def\GWs{{gravitational waves\;}}

\def\?{{\quad\gb{\fbox{\texttt{?}}\;}}\quad}

\def\p{{\partial}}

\def\Rarrow{{\quad\Rightarrow\quad}}

\def\p{\partial}

\def \p{{\partial}}

\def\benu{\begin{enumerate}}
\def\eenu{\end{enumerate}}
\def\bitem{\begin{itemize}}
\def\eitem{\end{itemize}}

\newcommand{\const}{\mathop{\rm const.}\nolimits}
\newcommand{\half }{\smallover{1}/{2}}

\def\smallover#1/#2{\hbox{$\textstyle\frac{#1}{#2}$}} %
\def\smallcirc{{\raise 0.5pt \hbox{$\scriptstyle\circ$}}}
\def\cabove(#1){\stackrel{\smallcirc}{#1}}
\def\ccabove(#1){\,\stackrel{\smallcirc\smallcirc}{#1}\,}
\def\cccabove(#1){\stackrel{\,\smallcirc\smallcirc\smallcirc}{#1}\,}
\def\2{{\smallover1/2}}

\def\cA{{\cal A}}
\def\boxit#1{
\vbox{\hrule\hbox{\vrule\kern4pt
\vbox{\kern5pt#1\kern5pt}\kern4pt\vrule}\hrule}
} 

\newcommand{\bigbox}[1]{\fbox{%
\rule[-20pt]{0pt}{45pt}$\;\;\displaystyle{#1}\;\;$}
}
\newcommand{\medbox}[1]{\fbox{%
\rule[-10pt]{0pt}{25pt}$\;\;\displaystyle{#1}\;\;$}%
}

\let\ssection=\section
\renewcommand{\section}
{\setcounter{equation}{0}\ssection}

\def\besub{\begin{subequations}}
\def\esub{\end{subequations}}

\begin{document}

\preprint{\texttt{arXiv:2112.09589v4 [gr-qc]}}

\title{Gravitational Waves and Conformal Time Transformations \\
}

\author{
P.-M. Zhang$^{1}$\footnote{mailto:zhangpm5@mail.sysu.edu.cn}, Q.-L. Zhao$^{1}$\footnote{mailto: zhaoqliang@mail2.sysu.edu.cn},
 and
P.~A. Horvathy$^{2}$\footnote{mailto:horvathy@lmpt.univ-tours.fr}
}

\affiliation{
${}^1$ School of Physics and Astronomy, Sun Yat-sen University, Zhuhai, (China)
\\
${}^2$  Institut Denis Poisson CNRS/UMR 7013 - Universit\'e de Tours - Universit\'e d'Orl\'eans Parc de Grandmont, 37200, Tours, (France)
\\
}

\date{\today}

\pacs{
04.20.-q  Classical general relativity;\\
}

\begin{abstract}

Recent interest in the ``memory effect'' prompted  us to revisit the relation of \GWs  to oscillators.
50 years ago Niederer [1] found that an isotropic harmonic oscillator with a constant frequency can be mapped onto a free particle. Later Takagi [2] has shown that ``time-dependent scaling'' extends the oscillator versus free particle correspondence to  a time-dependent frequency when the scale factor satisfies a Sturm-Liouville equation. More recently  Gibbons [3] pointed out that time redefinition is conveniently studied in terms of the Schwarzian derivative. The oscillator versus free particle  correspondence ``Eisenhart-Duval lifts'' to a conformal transformation between Bargmann spaces [4-7]. These methods are extended to spacetimes which are not conformally flat and have a time-dependent profile, and can then be applied to the geodesic motion in a plane gravitational wave.
\\

Annals Phys. \textbf{440} (2022), 168833
doi:10.1016/j.aop.2022.168833
[arXiv:2112.09589 [gr-qc]].

\medskip
\noindent{Key words: plane gravitational waves, geodesic motion, conformal time redefinition, anisotropic and time-dependent oscillators, Sturm-Liouville equation.}

\end{abstract}

\maketitle

\tableofcontents

\section{Introduction: gravitational waves and the Niederer transformation}\label{Intro}

Recent interest in the ``memory effect'' for \GWs   attracted renewed attention in oscillators and their relation to free particles \cite{ZDGHMemory,LukashII}.

The story starts with the seminal discovery of Niederer \cite{Niederer73} who found that the space-time transformation
\beq
N : (T,\bX) \to (t,\bx)~: \qquad
T=\tan t\,,
\qquad
\bX=\frac{\bx}{\cos t}\,,
\label{NiederTX}
\eeq
referred to as the \emph{Niederer map}
carries the Schr\"odinger equation  for a free  non-relativistic particle of unit mass with coordinates $\bX,T$ into that of an isotropic harmonic oscillator with  coordinates $\bx,t$ and constant frequency
$\omega_0=1$  \cite{Niederer73}.
The oscillator versus free particle correspondence has been studied subsequently by many people, see \cite{BurdetOsci,JI85,DBKP,DGH91,Takagi,DHP2,PerrinBD,GWG_Schwarz,Andr18,Andr2014,ZHAGK,AndrPrenc,Galajinsky,Inzunza,ZZH,Masterov,Silagadze, Curtright,Arnold,Aldaya} for a (very) incomplete list. Most of these studies concern the isotropic and time-independent case, though.
The Niederer transformation and its subsequent generalizations mentioned above fit into the general framework of Arnold transformations \cite{Arnold,Aldaya,CDGH}.

The relation between non-relativistic physics and \GWs was established in \cite{Eisenhart,DBKP,DGH91}.
The vacuum Einstein equations imply however that exact plane \GWs are \emph{anisotropic} and their space-times are therefore \emph{not conformally flat}~: threfore they can  \emph{not be mapped to free particles}.
The  gravitational plane wave considered  in \cite{Brdicka} is, for example, manifestly anisotropic;  the one proposed by Lukash \cite{LukashJETP75} in anisotropic cosmology  has, in addition, a complicated time-dependent profile. See \cite{Ehlers,LukashI,LukashII,exactsol} for details.

In this paper we study (possibly) anisotropic  oscillators with a (possibly) time-dependent  frequency, applied to \GWs \footnote{A ``bon mot'' attributed to S. Coleman cited in \cite{Curtright} says that ``The career of a young theoretical physicist consists of treating the harmonic oscillator in ever-increasing levels of abstraction.''}. We concentrate on classical aspects;  quantum aspects have (and will again) be studied elsewhere.

 Our clue is that the oscillator-versus-free particle  correspondence fits into the broader ``conformal'' context advocated by Gibbons \cite{GWG_Schwarz}, which will be our guiding thread throughout much of our investigations. Certain
\GW space-times can sometimes be mapped onto each other conformally; then our main result says that our framework can accommodate  \emph{anisotropic systems with time-dependent frequency and thus can be applied to \GWs}, see sect.\ref{GWsec}.

Our conventions are: Symbols written in boldface  as $\bx, \bX$ etc  refer typically to $d > 1$ space dimensions. In  $d=1$  the boldface is dropped and we write $x, X$. $V(\bx,t)$ written in generic coordinates $(\bx,t)$ is the potential we start with which is carried into $\Upsilon(\bxi,\tau)$ by time redefinition $t \to \tau$. $(\bxi,\tau)\leadsto (\bX,T)$ are in particular coordinates for a free system, $\Upsilon(\bX,T)\equiv0$. $t \to \theta$ is another time redefinition to a non-free system of the type discussed in subsec. \ref{thetasec}.

\section{Conformal redefinition of time}\label{Gibbsec}

In \cite{GWG_Schwarz} Gibbons proposed to consider a  redefinition $t \to \tau$ of time, implemented as,
\beq
t=t(\tau),
\quad
\bx = \big(\cabove(t)\big)^{1/2}\,\bxi,
\where
\cabove(\{\,\cdot\,\})= d/d\tau.
\label{txtauxi}
\eeq
This takes the {usual Hamiltonian action  of the oscillator,
${\cal A}=\displaystyle\int\! Ldt$, into
\beq
\int\Big\{\half \,
{\cabove(\bxi)}^2-\Upsilon(\bxi,\tau)\Big\}\,d\tau\,,
\label{xitauaction}
\eeq
where the potential
\beq
\Upsilon(\bxi,\tau)=
\, \cabove(t)V\left(\big(\cabove(t)\big)^{1/2}\,\bxi,\tau\right)
+\smallover{1}/{4}\calS_{\tau}(t) %
\,\bxi^2\,
\label{Upsipot}
\eeq
includes an additional isotropic oscillator term with a possibly time dependent frequency.
The 2nd term here is proportional to the \emph{Schwarzian derivative} of $t(\tau)$ w.r.t.  $\tau$ that we shall call henceforth ``fake time'',
\beq
\calS_{\tau}(t)
=\left(\frac{\ccabove(t)}{\cabove(t)}\right)^{\cabove()}
-\frac{1}{2}\left(\frac{\ccabove(t)}{\cabove(t)}\right)^2
=\frac{\cccabove(t)}{\cabove(t)}
-\frac{3}{2}\left(\frac{\ccabove(t)}{\cabove(t)}\right)^2\,.
\label{ttauS}
\eeq

We emphasise that $\tau$  can be chosen as we wish (up to regularity requirements). Choosing, for example \footnote{$\arctan$ is one of the branches of a multivalued function; considering all branches allows us to recover the Maslov correction \cite{BurdetOsci,ZZH}.},
\beq
t = \arctan T   \Rarrow    \bx = \frac{\bX}{\sqrt{1+T^2}}
\label{invNied}
\eeq
(where we have changed our notation, $\tau \leadsto T$ and $\bxi \leadsto \bX$), we find
$
\calS_{T}(t)=  - \frac{2}{(1+T^2)^2}\,.
$
Starting with  $V(\bx)= \half \bx^2$ the two terms in \eqref{Upsipot} then
cancel~: \eqref{invNied} \emph{maps the unit-frequency oscillator with coordinates $(\bx,t)$ to a free particle} with coordinates $(\bX,T)$.
\eqref{invNied} is the inverse of  the \emph{Niederer map} $N$ in \eqref{NiederTX}.
Reversing the roles in \eqref{txtauxi}-\eqref{Upsipot},
\beq
\tau =\tau(t), \quad \bxi=\dot{\tau}^{1/2}\bx,
\quad
V(\bx,t)=
\, \dot{\tau}\, \Upsilon\left({\dot{\tau}}^{1/2}\,x,t\right)
+\smallover{1}/{4}\calS_{t}(\tau) \,\bx^2\,,
\label{invcoordpot}
\eeq
where $\dot{\{\,\cdot\,\}}=d/dt$.
For $\Upsilon(\bxi,\tau)=\half \baromega^2 \,\bxi^2$
in particular (where $\baromega^2(\tau)$ can be positive or negative), \eqref{invcoordpot} yields
\beq\medbox{
V(\bx,t)=\half\omega^2\bx^2
\where
\omega^2=
\dot{\tau}^2\baromega^2+\half\calS_{t}(\tau)\,
}
\label{VfromUp}
\eeq
or conversely,
\beq\medbox{
\Upsilon(\bxi,\tau)=\half\baromega^2(\tau)\,\bxi^2
\where
\baromega^2(\tau) =
(\cabove(t))^2\,\omega^2(t)+\half\calS_{\tau}(t)\,.
}
\label{UpfromV}
\eeq
The result is yet another oscillator  with a shifted frequency.

These formulae work when the oscillator is isotropic, but \emph{fail} when it is anisotropic, see \eqref{twotau} in sec.\ref{ExpOex}.

Starting with a free particle, $\Upsilon \equiv 0$, \eqref{VfromUp} yields a generally time-dependent  isotropic oscillator,
\beq
V(\bx,t)= \smallover{1}/{4}\calS_{t}(\tau)\, \bx^2\,.
\label{FlatSchwpot}
\eeq
An oscillator with a given frequency
$\omega(t)$ can thus be obtained by a conformal time redefinition when $t\to\tau$ solves the 3rd order non-linear equation
\beq
\calS_{t}(\tau)\equiv \frac{\dddot{\tau}}{\dot{\tau}}-\frac{3}{2} \Big(\frac{\ddot{\tau}}{\dot{\tau}}\Big)^2 = 2\omega^2(t)\,.
\label{omegaosc}
\eeq
For the Niederer map \eqref{NiederTX}
$ \calS_t(T)=2,$
and we recover our oscillator with unit frequency $\omega_0=1$. See also \cite{Masterov}.

We stress that the ``fake time'' $\tau$ can be chosen freely. Choosing, for example, $\tau = \arcsin t$ \cite{Curtright} is allowed; however \eqref{FlatSchwpot} then yields an oscillator  with a complicated time-dependent profile $\calS_t(\tau) =\frac{t^2+2}{2(t^2-1)^2}$. Another (somewhat unorthodox) choice may instead eliminate the kinetic term and promoting the Schwarzian to a (higher-order) Lagrangian \cite{Masterov}.

The oscillator potential with a shifted frequency induced by a conformal redefinition, \eqref{VfromUp} or \eqref{UpfromV}, is essentially the one proposed by \JI \,\cite{JI85}, who studied $\baromega=\const$. See also \cite{ZZH}.
The Schwarzian plays an important role in the coupling-constant morphoses see refs.\cite{Mertens,Plyushchay:2016lid}.
Our ``Schwarzian'' calculation here is also consistent  with eqn. \# (36) in \cite{Silagadze}.

Non-flat and. non-flat examples with particular relevance for general relativity will be considered in secs. \ref{ExpOex} and \ref{GWsec}.

\subsection{Time-dependent rescaling}\label{TakagiSec}

A variation of the Niederer construction was proposed Takagi in his ``co-moving framework'' \cite{Takagi}. He first dilates the position with a time-dependent scale factor and then changes the time accordingly,
\beq
\bxi = \frac{\bx}{a(t)}, \qquad \tau =\int^t\! \frac{dt}{a^2(t)}\,
\where
a = a(t)\,.
\label{Tredef}
\eeq

This shifts the potential according to his eqn \#(9),
\beq
V(\bx,t) = \dfrac 1 2 \omega^2(t) \bx^2 \; \to \; \Upsilon(\bxi,\tau) = \half a^3\Big(\ddot{a}+\omega^2(t)a\Big)\,\bxi^2\,.
\label{ashiftedpot}
\eeq
Thus if the scale factor $a(t)$ satisfies the \emph{Sturm-Liouville equation} then the potential is eliminated,
\beq\medbox{
\ddot{a}+\omega^2(t)a =0\, \Rarrow \Upsilon \equiv 0\,}
\label{TSL}
\eeq
cf. \# (10) of \cite{Takagi}  leaving us with a \emph{free particle}. See also \cite{BurdetOsci} for another approach.

When $\omega = \omega_0=\const$ ($\omega_0=1$ for example) then choosing
$
a(t) = \cos  t
$
we recover Niederer's choice \eqref{NiederTX}.
Eqn. \eqref{TSL} implies that the scale factor is not unique: choosing instead $a(t)=\sin t$, for example, yields
$
T = -\cot t$ and
$\bX = \frac{\bx}{\sin t}\,,
$
which carries $V(\bx)=\half \bx^2$ also to the free case.
The difference between these choices concerns the singularities: while Niederer's choice works where $\cos \neq0$ e.g. in the half-periods $(-\pi/2+k\pi, \pi/2+k\pi)$ ($k$ is an integer), the sine-choice works in the half-periods $(k\pi,(k+1)\pi)$. Subtleties related to domains were studied e.g. in \cite{BurdetOsci,ZZH}.

For the sake of comparison we consider $t= t(\tau)$  and  recall that the potential in \eqref{FlatSchwpot} matches a given oscillator when \eqref{omegaosc} is satisfied.
Putting here, consistently with \eqref{Tredef},
\beq
\dot{\tau}= a^{-2}(t)
\Rarrow
\calS_{t}(\tau)=-2\ddot{a}/{a}
\label{taua}
\eeq
shows that \eqref{omegaosc} requires precisely the S-L equation \eqref{TSL}.
In terms of the Schwarzian cocycles \cite{Duval14}
\beq
{\cA}\equiv{\cA}_{t}(\tau) =
\frac{\ddot{\tau}}{\dot{\tau}}
=-2\,\frac{\dot{a}}{a}\,,
\qquad
\calS_{t}(\tau)=\dot{\cA}-\frac{1}{2}\big({\cA}\big)^{2},
\label{affincoc}
\eeq
and thus \eqref{omegaosc} requires that ${\cA}$ satisfy the first-order non-linear Riccati equation  \cite{BurdetOsci},
\beq
\dot{{\cA}}-\frac{1}{2}{\cA}^{2}-2\omega^{2}(t)=0\,.
\label{Riccatieq}
\eeq

The combined Takagi-Gibbons approaches goes beyond the constant-frequency assumption and hints also at the  $\omega\neq\const$ case --- at least theoretically. The practical difficulty is to solve either of the equations \eqref{TSL}, \eqref{omegaosc} or \eqref{Riccatieq} for a given time-dependent frequency $\omega(t)$, which can be done analytically only in exceptional cases; otherwise, one has to resort  to numerical work.
The time-dependent frequency  $\omega(t)=\cos t$  leads, for example, to the Mathieu equation \cite{ZZH}. A new, non-trivial but instructive  illustration is presented in sec.\ref{ExpOex}.

Takagi's co-moving framework is clearly equivalent to that of Gibbons we follow~: deriving $\tau$ in \eqref{Tredef}
yields $\dot{\tau}= a^{-2}(t)$ and then the first relation reduces to the one in \eqref{txtauxi}.
Takagi's cosmological applications  (co-moving coordinates, Hubble law, Friedmann-Robertson-Walker universe, etc.) justify his ``co-moving frame" terminology as alternative to ``fake time'' and ``conformal redefinition" we use here.
\goodbreak

\section{The Eisenhart-Duval lift}\label{Bargmannsec}

Further insight can be gained by ``Eisenhart-Duval (E-D) lifting'' the system to one dimension higher,   called, in the present context, a ``Bargmann space'' \cite{Eisenhart,DBKP,DGH91}.
 The latter is a $d+1+1$ dimensional manifold endowed with a Lorentz metric considered a long time ago by Brinkmann \cite{Brinkmann}, whose general form is \footnote{Here $\bx = (x^i)$. In what follows we focus our attention to $d=2$ so $i=1,2$.},
\begin{equation}
ds_{B}^2=
 g_{ij}(x,t)dx^{i}dx^{j}+2dtds-2V(\bx,t)dt^{2}\,.
\label{genBmetric}
\end{equation}
It carries a covariantly constant null Killing vector $\p_s$. Then we have~:

\vskip-2mm
\kikezd{Theorem 1 \cite{DBKP,DGH91}}~: \textit{Factoring out the foliation generated by $\p_s$ yields  a non-relativistic space-time in $d+1$ dimensions.
 Moreover, the null geodesics of the Bargmann metric \eqref{genBmetric}
 project to ordinary space-time consistently with Newton's equations for $V$ as scalar potential. Conversely, a solution $(\bgamma(t),t)$  of the non-relativistic equations of motion lifts to Bargmann space as a null geodesic, \vspace{-1mm}
\beq
\big(\bgamma(t),t,s(t)\big),
\quad
s(t)=s_0-\int^t\!\!L\big(\bgamma(t),t\big)dt
\label{Blift}
\eeq \vspace{-1mm}
where $s_0=\const$ is an arbitrary initial value and $\displaystyle\int\!\! L\big(\bgamma(t),t\big)dt$ is the classical Hamiltonian action calculated along $(\bgamma(t),t)$.
}

A case of particular interest is that of \emph{exact plane gravitational  waves} which correspond to potentials of the form
\beq
V(\bx,u)=-\half K_{ij}(u) x^i x^j,\qquad
ds^2=d\bx^2 + 2 du dv + K_{ij}(u) x^i x^j du^2\,,
\label{BKmetric}
\eeq
where $K_{ij}(u)$ is a symmetric and traceless $2\times2$ matrix \cite{exactsol,LukashJETP75,Ehlers},
and $\bx=\big(x^1,x^2\big)$.
The vacuum Einstein equations are satisfied for an arbitrary profile $K_{ij}(u)$ which is in fact decomposed into $+$ and $\times$ polarization states,
\beq
K_{ij}(u){x^i}{x^j}=
\half{h}_{+}(u)\big((x^1)^2-(x^{2})^2\big)\,+\,{h}_{\times}(u)\,x^1x^2\,.
\label{BKprofile}
\eeq
We restrict our attention to metrics of the form \eqref{BKmetric}-\eqref{BKprofile} in what follows.

$\bullet$ A free particle is described by Minkowski space whose metric is $d\bX^2+2dTdS$.

$\bullet$ An isotropic oscillator in $d$ dimensions is described in turn by
the d+2 dimensional Bargmann manifold with metric
\begin{equation}
ds^2_{osc}=d\bx^2+2dtds-\omega(t)^2\bx^2dt^{2}\,
\label{oscimetric}
\end{equation}
which is not a vacuum Einstein solution.
For $\omega(t)=\omega_0=\const$ the Niederer correspondence  \eqref{NiederTX} of $d+1$ dimensional non-relativistic spacetime lifts to one between $d+2$ dimensional Minkowski and the oscillator Bargmann spaces \cite{BurdetOsci,DBKP},
\beq
T=\frac{\tan\omega_0 t}{\omega_0}\,,
\qquad
\bX=\frac{\bx}{\cos\omega_0 t}\,,
\qquad
S=s-{\half}\bx^2\omega_0 \tan\omega_0 t\,.
\label{NiedEisen}
\eeq
The mapping $(\bx,t,s) \to (\bX,T,S)$ in \eqref{NiedEisen} is conformal with conformal factor
$\Omega^2=\cos^2\omega_0 t$\,.
 More generally \cite{Duval14}, expressing the Schwarzian in terms of the {affine cocycle} $\calA_{\tau}(t)={\ccabove(t)}/{\cabove(t)}$
as in \eqref{affincoc} and completing \eqref{txtauxi} by
\beq
\sigma = s + \smallover1/4%
\calA_{\tau}(t)\,{\bxi}^2
\quad\Leftrightarrow\quad
s = \sigma +\frac{1}{4}\calA_{t}(\tau)
\mathbf{x}^2\,
\label{ssigmalift}
\eeq
implies
\beq
ds^2_B = \,
 \cabove(t) \underbrace{\Bigl\{d{\bxi}^2+ 2 d\tau d\sigma
 - 2\Upsilon(\bxi, \tau)\, d\tau^2\Bigr\}}_{d\Sigma^2}\,.
\label{BConfRescale}
\eeq
Thus the Bargmann metrics \eqref{genBmetric} and  $d\Sigma^2$
inside the brace are conformal but are not isometric  when $\tau\neq{t}$.

\subsection{Conformally flat Bargmann manifolds}\label{ConfflatB}

Quadratic potentials  which can be mapped to a free particle by a conformal time rescaling are necessarily isotropic and could (in principle), be found by solving  \eqref{omegaosc} for $\tau(t)$.
An alternative way is to find all Bargmann spacetimes which are \emph{Bargmann-conformally flat}, by which we mean that (i) they can be mapped to free (Minkowski) space in such a way that (ii) their respective covariantly constant vectors $\p_S$ and $\p_s$ are intertwined.

The importance of conformal flatness is underlined because it implies Schr\"odinger symmetry \cite{Niederer73,DBKP,DGH91}). Conformal flatness can be determined by geometric methods \cite{DHP2,Silagadze}.
\begin{itemize}
\item
In $d=1$ i.e. 3 total dimensions, the conformal properties are fixed by the \emph{Cotton tensor},
\beq
C_{\mu\nu\lambda}=
\nabla_{\lambda}R_{\mu\nu}-\nabla_{\nu}R_{\mu\lambda}+
\smallover1/4\big(g_{\mu\lambda}\nabla_{\nu}R-g_{\mu\nu}\nabla_{\lambda}R\big)\,.
\label{Cottontensor}
\eeq
Requiring it to vanish  implies
$
\frac{\p^3V(x,t)}{\p x^3} = 0,
$
 allowing us to deduce the most general potential,
\beq
V(x,t)
= \smallover1/4 \calS_t(T) x^2 + F(t)x + K(t),
\label{1flatpot}
\eeq
where $S_t(T)$ is the Schwarzian derivative of the arbitrary time redefinition $t \to T$. (cf. eqns. (\# (23) and \# (36) of ref \cite{Silagadze}).
Physically, we have a $1$d oscillator combined with a uniform force $F$.

\item
B-conformal flatness has been investigated earlier \cite{DHP2}. In $D=d+2= 4$ total dimensions conformal flatness is guaranteed  by the vanishing of the \emph{conformal Weyl tensor}
\beq
C^{\mu\nu}_{\ \ \rho\sigma}
=
R^{\mu\nu}_{\ \ \rho\sigma}
-
\smallover 4/{D-2}
 \,\delta^{[\mu}_{\ [\rho}\,R^{\nu]}_{\ \sigma]}
+
\smallover 2/{(D-1)(D-2)}\,
\delta^{[\mu}_{\ [\rho}\,\delta^{\nu]}_{\ \sigma]}\,R\,.
\label{CWeyl1}
\eeq
 Referring to  \cite{PerrinBD,DHP2}
 for details, we just mention that for a pp wave the conformal Weyl tensor takes the
form
\beq
C^{\mu\nu}_{\ \ \rho\sigma}=
R^{\mu\nu}_{\ \ \rho\sigma}
-
2\,
\varrho\,\delta^{[\mu}_{\ [\rho}\,\xi^{\nu]}\xi^{ }_{\sigma]}
\label{ppWeyl}
\eeq
and  B-conformal flatness requires  \vspace{-3mm}
\beq\vspace{-1mm}
C_{\mu\nu\rho\sigma}\,\xi^\mu=0\,,
\label{BW0}\vspace{-2mm}
\eeq
(where $(\xi^\mu)=\p_s$). From this
one can conclude~:

\vspace{-3mm}
\kikezd{Theorem 2} \cite{DHP2}. \textit{The most general B(argmann)-conformally flat Bargmann metric in $D=4$ total dimensions is,
\beq
d\bx^2+2dt\big[ds+{\bA}\cdot d{\bx}\big]-2V\,dt^2
\label{Bvecpot}
\eeq\vskip-3mm
such that\vspace{-3mm}
\besub
\begin{align}
&{A}_i(\bx,t)=\2\epsilon_{ij}{B}(t)x^j+a_i,
\qquad
\vec{\nabla}\times\vec{a}=0,
\qquad\partial_t\vec{a}=0,
\label{Bfield}
\\[6pt]
&V({\bx},t)=\2 C(t)\bx^2+{\bF}(t)\cdot{\bx}+K(t)
\label{Vfield}
\end{align}
\label{Wflat}
\esub
where $\bx=(x^i),\, {\bA}=(A_i),\,\vec{a}$ and ${B}$ refer to the transverse plane.}
\goodbreak

From the Bargmann point of view, the  metric  \eqref{Bvecpot}-\eqref{Wflat} describes
\bequ
\item
a \emph{uniform magnetic field} ${B}(t)$,
\item
an \emph{attractive  or repulsive} [$C(t)=\pm\omega^2(t)$]
\emph{\underline{isotropic} oscillator}

\item
a uniform force field ${\bF}(t)$ \eenu \goodbreak
which may all depend arbitrarily on time.
It also includes a curl-free
vector potential~$\vec{a}({\bx})$ that can be gauged away if the
transverse space is simply connected.
 If, however, space is not simply connected,  an external \emph{Aharonov-Bohm-type vector potential} can also be included \cite{Jackiw90}.
\end{itemize}

The Bargmann metric of a constant force field ${\bF}$, for example,
\beq
d\bx^{2}+2dtds-2\bF\cdot\bx dt^{2},
\label{Fmetric}
\eeq
can be brought to the free form $d\bX^2+2dTdS$ by the {isometry at fixed time} \cite{DHP2},
\beq
T=t,
\quad
{\bX}=\bx+\2{\bF}\,t^2,
\quad
S=s-{\bF}\cdot\bx\,t-\smallover1/6{\bF}^2t^3 \,.
\eeq
A uniform magnetic field and an oscillator potential  \eqref{Bfield} - \eqref{Vfield} could actually be converted into each other.
For  $B=\const$ and again for fixed time,
$(t,\bx,s) \to (\tau,\bxi,\sigma)$,
\beq
\tau=t,
\quad
\xi^i=x^i\cos\omega t+\epsilon^i_{\,j}x^j\sin\omega t,
\quad
\sigma=s
\where \omega = \dfrac{B}{2}
\label{magtoosci}
\eeq
 takes the constant-${B}$-metric \cite{DHP2} \eqref{Bvecpot}  with $V=0$,
\beq
d\bx^2+2dt\Big(ds-\half{B} \epsilon^i_{\,j}x^idx^j\Big)
\label{constBmetric}
\eeq
isometrically into the oscillator metric
$
d\bxi^{2}+2d{\tau}d\sigma-\omega^{2}\bxi^{2}d\tau^{2}\,.
$
Following \eqref{magtoosci} by the (inverse) Niederer map,
\beq
(t,\bx,s) \to (\tau,\bxi,\sigma)\to (T, \bX, S)
\eeq
transforms the constant-B metric to a free particle.

More generally, keeping $\omega$ and $B$ independent, the rotation \eqref{magtoosci} would yield a mixture of constant-magnetic and oscillator fields,
\beq
d{\bxi}^{2}+2d \tau d \sigma+2d \tau \left(\omega -\frac{B}{2}\right) \left(\xi^1d\xi^2 -\xi^2 d\xi^1\right) - \omega \left(- \omega +B\right) {\bxi}^{2}d\tau^{2}\,.
\label{magendo}
\eeq
In conclusion, the B-metrics which describe constant magnetic and isotropic oscillators are related by a change of coordinates.

The correspondence \eqref{BConfRescale} between Bargmann spaces can be further extended \cite{Duval14}. Starting with the Bargmann metric \eqref{Bvecpot} which includes a vector potential $\bA$, \eqref{txtauxi} completed with \eqref{ssigmalift}
extends \eqref{BConfRescale} into a B-conformal diffeomorphism $(\bxi,{\tau},{\sigma})\mapsto(\bx,t,s)$ which involves both the
 Schwarzian and the affine cocycles in \eqref{affincoc},
\besub
\begin{align}
ds_B^2 \,&=\, \cabove(t)\underbrace{
\Bigl\{d\bxi^2+2d\tau\big[d\sigma+\mathfrak{A}\cdot d{\bxi}\big]-2\Upsilon\,d\tau^2
\Bigr\}}_{d\Sigma^2}
\label{Bconfds2}
\\[4pt]
{\bA}& = \big(\cabove(t)\big)^{-1/2}\,\mathfrak{A}\,,
\qquad
V=\big(\cabove(t)\big)^{-1}
\left[\Upsilon-\smallover{1}/{4}\calS_t(\tau)\,\bxi^2+\half\,\calA\,\mathfrak{A}
 \cdot\bxi\right]\,.
\end{align}
\label{DuvalAV}
\esub\vspace{-10mm}

\section{Time-dependent frequency~: a toy example
}\label{ExpOex}

Now we illustrate our theory by a $1d$ oscillator with the time-dependent frequency
\beq
\omega(t)=e^{-t/2}\,,
\label{expofreq}
\eeq
where, to avoid infinite growth, we restrict our attention at $t>0$.
\goodbreak

First we \emph{carry the time-dependent system to a free one}. Following the Takagi approach, we search for a scale factor $a_{+}(t)$ which is (as we have seen in subsec.\ref{TakagiSec}) a solution of the Sturm-Liouville equation \eqref{TSL},
\beq
\ddot{a}_{+}+ e^{-t}a_{+} = 0\,
\label{expoSL}
\eeq
where a suffix $\{\,\cdot\,\}_{+}$ was added for later convenience.

For large $t$ the frequency  falls off exponentially and \eqref{expoSL} reduces to
$
\ddot{a}_{+}\sim 0.
$
Therefore $a(t)_+$ is  approximately a linear function, $a_{+}(t) \sim \alpha_1 + \alpha_2 t\,.
$
Eqn  \eqref{expoSL}  can actually be solved exactly by redefining the time,
$
t \to \theta =2e^{-t/2}\,.
$
 Denoting $\{\,\cdot\,\}^{\prime}=d/d\theta$,
\eqref{expoSL} becomes Bessel's equation of order zero,
\beq
{\theta}a_{+}^{\prime \prime}+a_{+}^{\prime} +{\theta}a_{+}=0\,
\Rarrow
a_{+}\left(t\right) =\alpha_{1}J_{0}\left(2e^{-t/2}\right) +\alpha
_{2}Y_{0}\left(2e^{-t/2}\right)\,,
\label{J0Y0eq}
\eeq
where the $\alpha_{1,2}$ are free constants and
 $J_{0}$ and $Y_{0}$ are the zero-order Bessel functions of the first and second kind, respectively, depicted in fig.\ref{J0Y0figs}i.
We shall return to ``$\theta$-time''
in sec. \ref{thetasec}.

Once we have found the desired scale factor which eliminates the potential, the ``fake time'' (we rename  $\tau \leadsto T$) is obtained by calculating the integral in \eqref{Tredef}, cf. fig.\ref{J0Y0figs}ii.
\begin{figure}
\hskip-2mm
\includegraphics[scale=0.31]{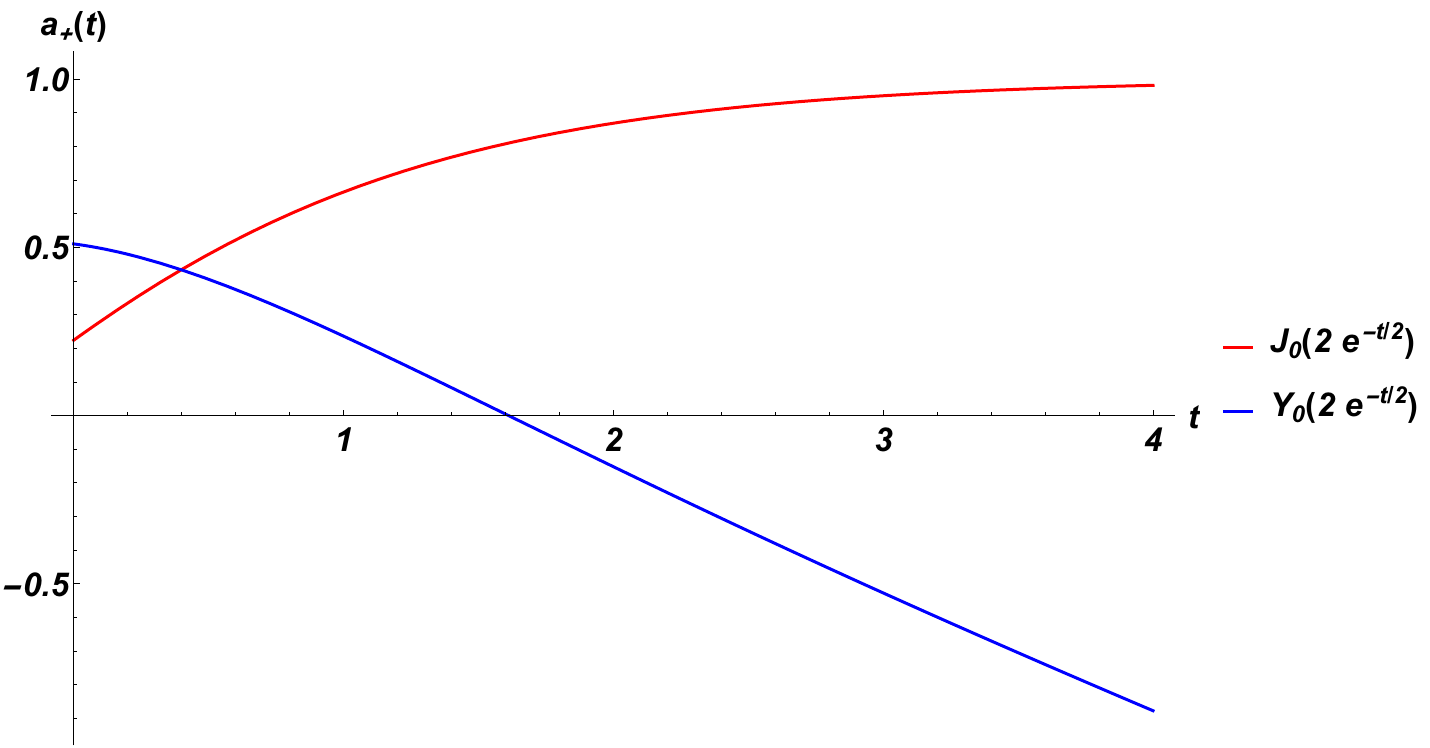}\,
\includegraphics[scale=0.33]{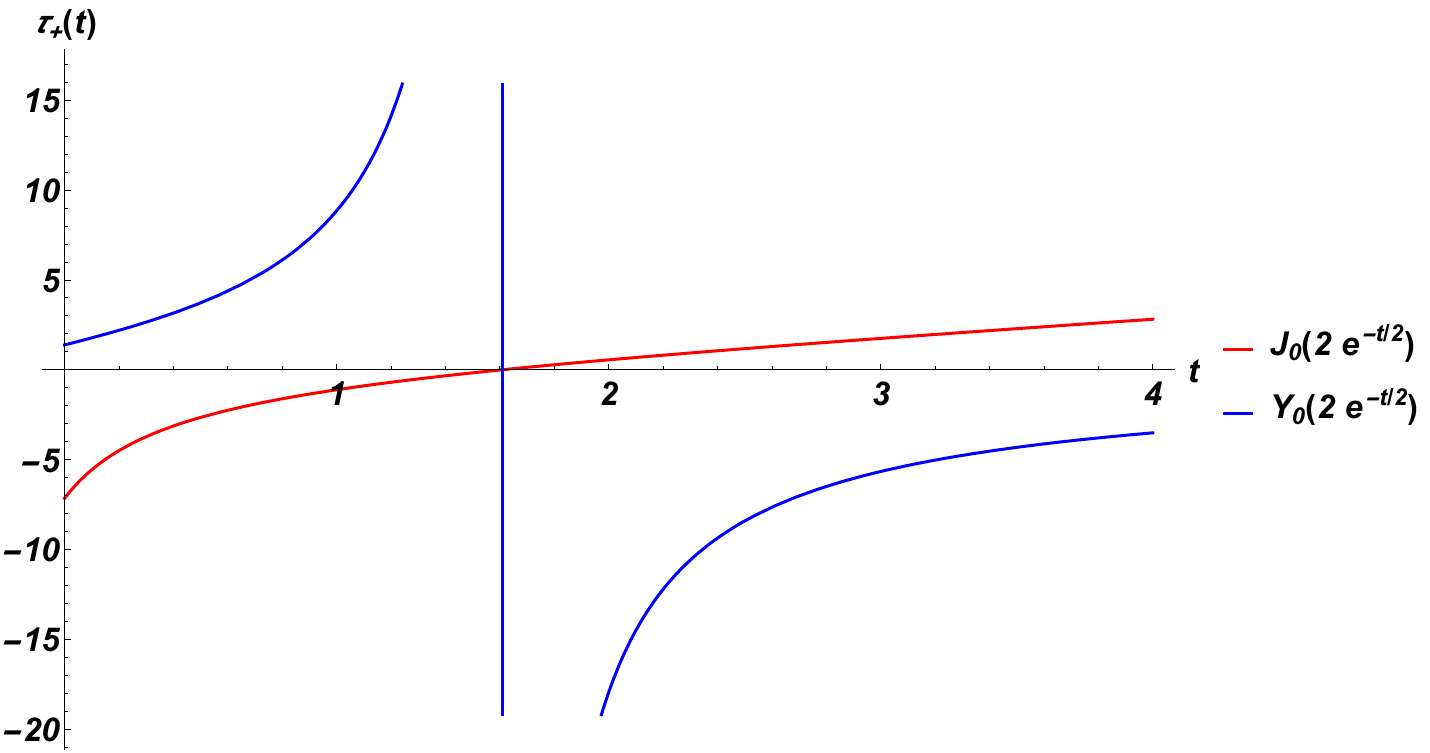}\\
\hskip-25mm
(i) \hskip75mm (ii)\\
\vskip-3mm
\caption{\small \textit{
(i) By \eqref{J0Y0eq} the Takagi scale factor $a(t)\equiv a_{+}(t)$ is a combination of Bessel functions of the first and second kind with $\theta=2e^{-t/2}$ in its argument. $\red{a(t)}$ (shown in \red{red}) is regular and behaves as $\red{a(t) \approx \const}$ for large $t$.
$\blue{a(t)}$ (shown in \blue{blue}) has instead a zero, and  $\blue{a(t) \approx -t}$ for large $t$.
 (ii) The fake time $\tau(t)$ can be found by numerical integration of $a^{-2}(t)$ in \eqref{Tredef}. \red{$\tau$} (shown in \red{red}) is regular and $\red{\tau \propto t}$ for large $t$.
\blue{$\tau$} (shown in \blue{blue}) is singular where the \blue{$a(t)$} is zero and \blue{$\tau$ $\propto -{t}^{-1}$} for large $t$.}
\label{J0Y0figs}
}
\end{figure}

Alternatively $a_{+}^{-2}$ and hence $\tau_+(t)=\int^t\!\!\frac{dt}{a_+^2(t)}$ can be calculated analytically,
\beq
a_{+}^{-2} = \left\{
\begin{array}{r}
\frac{1}{J^2_0(2e^{-t/2})}
\\[9pt]
\frac{1}{Y^2_0(2e^{-t/2})}
\end{array}
\right.
 \Rightarrow\;\;\;
\tau_{+}(t) = \left\{
\barraynb{llll}
- \frac{Y_0 (2 e^{-t/2})}{J_0(2e^{-t/2})}&\quad
&\text{\small for}\, &a_+=J_0(2e^{-t/2})
\\[12pt]
\;\;\, \frac{J_0 (2 e^{-t/2})}{Y_0 (2 e^{-t/2})}&\quad
&\text{\small for} \, &a_+=Y_0(2e^{-t/2})
\earraynb\right.\;.
\label{expofaketime}
\eeq
The analytic and numerical results are consistent.
The independent solutions $J_0$ and $Y_0$ of \eqref{expoSL} generate a two-parameter family whose elements transform the oscillator to the free form.
Equivalent results could be obtained by solving instead the Riccati equation \eqref{Riccatieq}.

To make contact with the approach of sec.\ref{Gibbsec} we start with the free case $\Upsilon\equiv0$ and work backwards. Then eqn. \eqref{FlatSchwpot} yields the oscillator potential with frequency square
$
-{\ddot{a}_+}/{a}_+
 =\omega^2(t)$, using that $a_+$ is a solution of \eqref{TSL}.

Having chosen a free particle versus oscillator
mapping  (with $\xi, \tau \leadsto X, T$), we pull back a free motion
$
X(T)= A + B T\,,\, A,B = \const
$
to get,  by \eqref{Tredef} and \eqref{J0Y0eq},
\beq
 x(t)= a_+(t)\,X(t) =
 \Big(\alpha_{1}J_{0}\left(2e^{-t/2}\right)
+
\alpha_{2}Y_{0}\left(2e^{-t/2}\right)\Big)\Big(A+BT(t)\Big)\,.
\label{importmotion}
\eeq
The motion in the oscillator
 potential $V(x,t)=\half e^{-t} x^2$ is given by the \emph{same} S-L eqn. \eqref{expoSL} with $a_+(t)$  replaced by $x(t)$, as it can be checked also directly, using the explicit form \eqref{importmotion}.

Instead of the attractive case  $\omega^2_{+} > 0$ in
 \eqref{expoSL} we can also consider the repulsive one,
\beq
\omega^2_{-}(t) = -e^{-t} < 0\,.
\label{repfreq}
\eeq
Then the \emph{same redefinition} $t \to \theta$  would yield instead the \emph{modified Bessel equation} of order zero, whose solutions are  combinations of the modified Bessel functions $I_{0}$ and $K_{0}$, respectively, depicted in fig.\ref{I0K0figs}i\,,
\beq
{\theta}a_{-}^{\prime \prime}+a_{-}^{\prime}-{\theta}a_{-}=0\,
\Rarrow
a_{-}(t) =\beta_{1}I_{0}\left(2e^{-t/2}\right)
+\beta_{2}K_{0}\left(2e^{-t/2}\right)\,,
\label{I0K0eq}
\eeq
($\beta_{1}, \beta_{2}=\const$).
Then
\begin{eqnarray}
a_{-}^{-2} = \left\{\begin{array}{r}
\frac{1}{I^2_0(2e^{-t/2})}
\\[9pt]
\frac{1}{K^2_0(2e^{-t/2})}
\end{array}
\right. \Rightarrow\;\;
\tau_{-}(t)= \left\{
\begin{array}{rll}
\;\frac{K_0(2e^{-t/2})}{I_0(2e^{-t/2})}
\ &\mathrm{for}
&a_-=I_0(2e^{-t/2})
\\[12pt]
-\frac{I_0(2e^{-t/2})}{K_0(2e^{-t/2})}
\ &\mathrm{for} \ &a_-=K_0(2e^{-t/2})
\end{array}
\right. \;. \quad
\label{reptau}
\end{eqnarray}
\begin{figure}
\hskip-3mm
\includegraphics[scale=0.32]{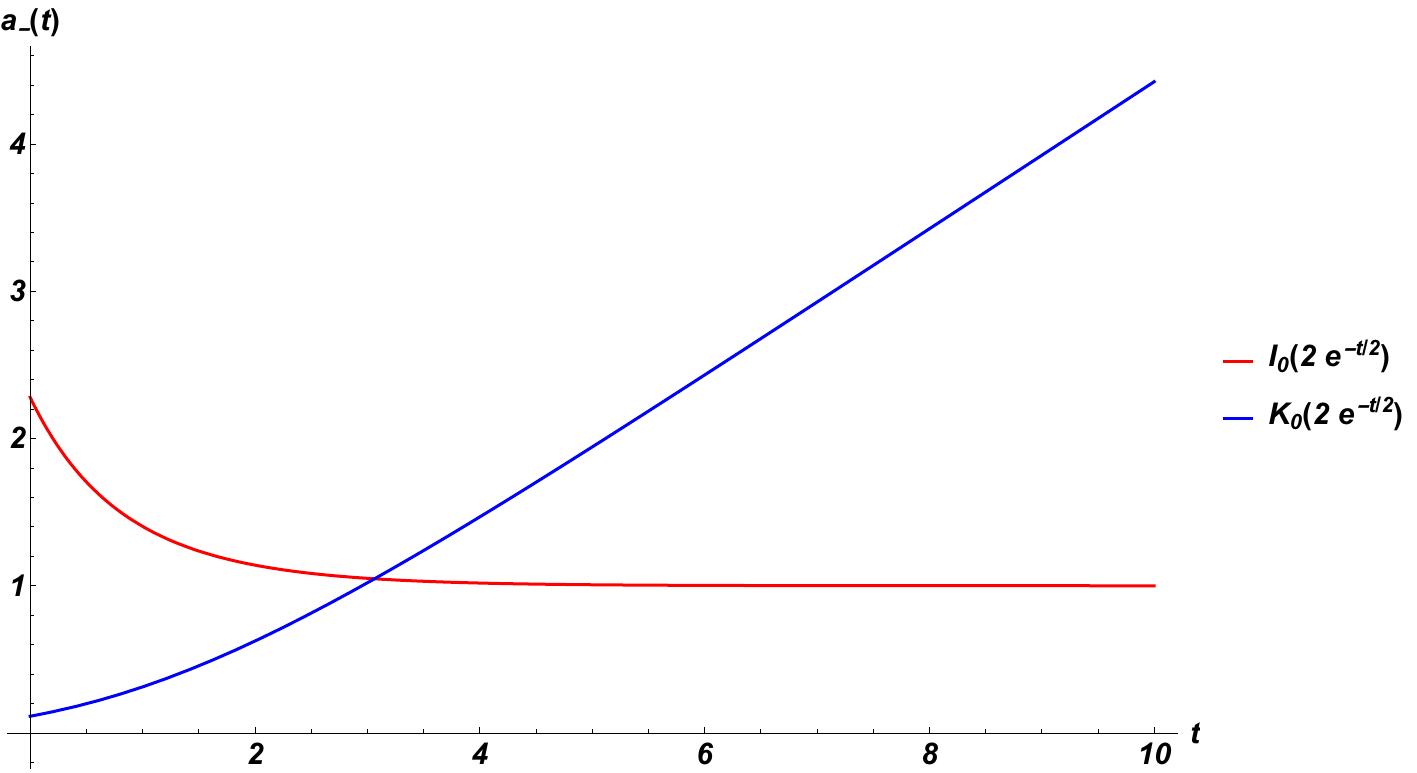}\,
\includegraphics[scale=0.33]{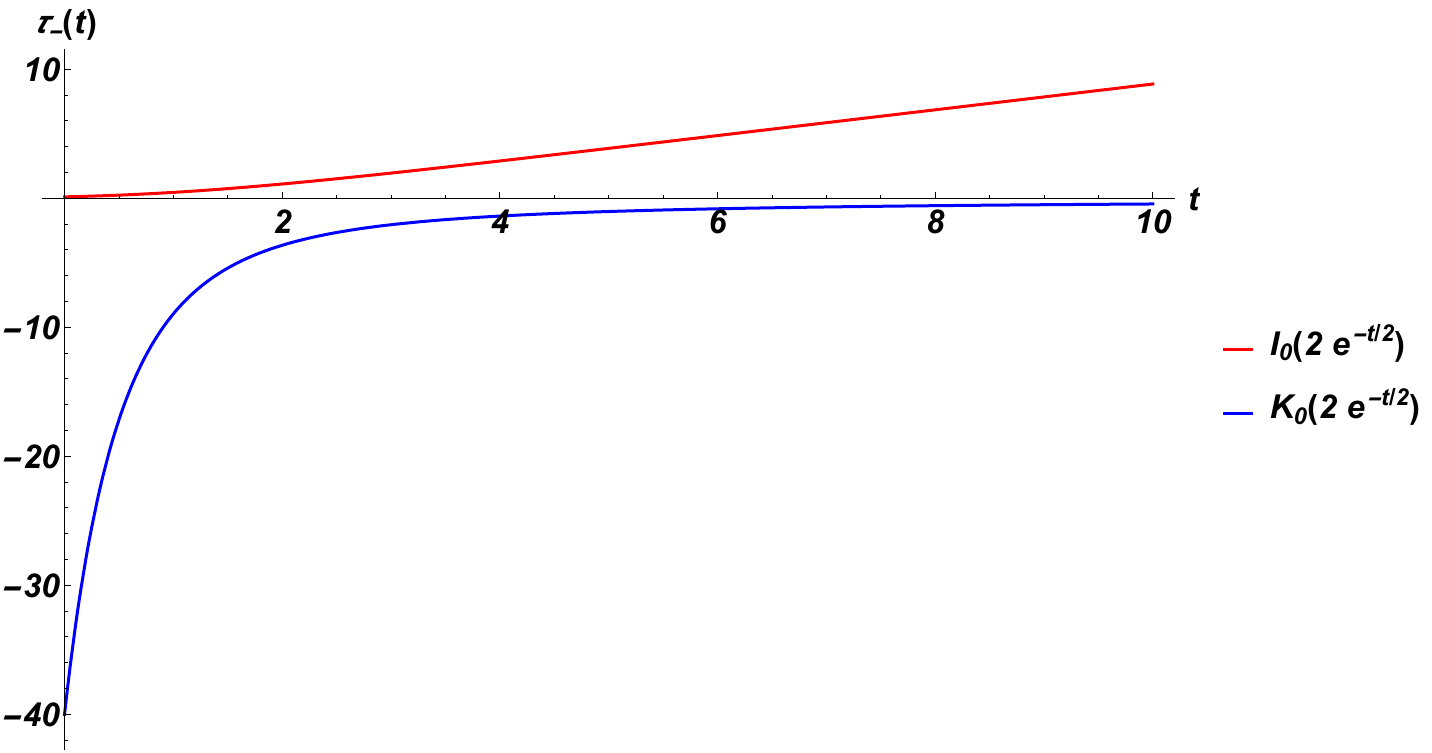}
\\
\hskip-20mm
(i) \hskip70mm (ii)\\
\vskip-3mm
\caption{\small \textit{(i) Scale factors $a_{-}(t)$ alias particle trajectories and (ii) fake times $\tau_-(t)$ for the repulsive potential \eqref{repfreq}, cf. \eqref{reptau}.
}
\label{I0K0figs}
}
\end{figure}
It is worth noting that $\cabove(\tau)_{\pm}\; > 0$ in both (attractive and repulsive) cases and therefore ${\big(\cabove(\tau)\big)}^{1/2}$ is real ; then the method of sec.\ref{Gibbsec} transforms them both (but separately), to their respective free forms. Let us underline, though, that the fake times in the attractive and repulsive cases,
\eqref{expofaketime} and \eqref{reptau}, respectively, are \emph{different},
\beq\medbox{
\tau_{+} (t) \neq \tau_{-} (t)\,.\;}
\label{twotau}
\eeq
We shall return to this point in sec.\ref{thetasec}, where  $\theta$ will be promoted to a new conformal time.

The  singularity of the ``blue $\tau$-time'' $\sim Y_0$ in fig.\ref{J0Y0figs}ii  highlights
 the caustic problem which is typical for oscillators and leads, at the quantum level, to the Maslov correction \cite{FeynmanHibbs,Schulman,KLBbook,Maslov, HFeynman,ZZH}. It  comes from the vanishing of the Bessel function in the denominator: the positive $t$-axes corresponds to the interval
$2\geq \theta=2e^{-t}>0$ where  $Y_0$ has one zero at $t \approx 1.6$. ($J_0$, $I_0$ and $K_0$ have no zeros).
\goodbreak

Physical examples which are relevant for the study of gravitational waves will be presented in sec.\ref{GWsec}.

\subsection{Another time redefinition~: $\theta$-time}\label{thetasec}

The combined Takagi-Gibbons approach  we followed so far carries \emph{conceptually} the original system in $d=1$ to a \emph{free} form for any scale factor $a(t)$, which is a solution of \eqref{expoSL}~\footnote{The method works also in $d>1$ when the oscillator is isotropic. Generalization to $d > 1$ is however problematic when the system is anisotropic, as will be discussed it in sec. \ref{GWsec}. }. Finding $a(t)$ may by difficult, though, ---  solving a S-L equation with time dependent frequency is always a challenging task.
In the toy model discussed above we could overcome the hurdle by switching to yet another ``time'' parameter,
\beq
t \to \theta = 2e^{-t/2} > 0\,,
\label{thetat}
\eeq
which converts the time-dependent S-L equation into one [\eqref{J0Y0eq} or \eqref{I0K0eq}] which, \emph{although not free and manifestly $\theta$-dependent}, we \emph{could} solve.

The redefinition \eqref{thetat} is thus a ``useful step'' --- but how does it fit into our framework~?  \emph{Could we view $t\to\theta$ as yet another ``conformal time redefinition"}, at the same footing as $t \to \tau$ in \eqref{txtauxi}~?

We first check that for $\omega^2(t)=e^{-t}$  the classical action is,
in terms of $\theta$,
\beq
\int\!L_{+}dt =
\int\!\left\{\half{\dot{x}_{+}}^2-\half{\omega_+^2}x_{+}^2\right\}dt =
- \half\int\!\theta\left\{\half{x_{+}'}^2-\half x_{+}^2\right\}d\theta\,,
\label{thetaact}
\eeq
where the prime means $d/d\theta$. The variation of this action reproduces the Bessel equation and its solutions \eqref{J0Y0eq}.
In terms of the new ``time'' parameter $\theta$ the oscillator metric \eqref{oscimetric} is written, accordingly,
\beq
ds_{osc+}^2 =
\frac{1}{\theta}\left(\theta dx_{+}^{2}-4d\theta ds_{+}-\theta x_{+}^{2}d\theta^{2}\right)\,,
\label{thetametric}
\eeq
whose geodesic equation is \eqref{J0Y0eq}.
Moreover, completing \eqref{thetat} by setting
\beq
x_+ = {\theta}^{-1/2}\,\zeta_+
\aand
\sigma_{+} = - 2s_{+} - \frac{\zeta_+^2}{4\theta}\,,
\label{zetathetasigma+}
\eeq
the oscillator metric can be expressed as
\beq
ds_{osc+}^2=\frac{1}{\theta}\,d\Sigma_+^2
\where
d\Sigma_+^2=d\zeta_+^{2}+2d\theta d\sigma_{+}-\Big(1+\frac{1}{4\theta^{2}}\Big) \zeta_+^{2}\,d\theta^{2}\,.
\label{zetathetasigmamet}
\eeq
$ds_{osc+}^2$ and $d\Sigma^2_+$ are thus conformal Bargmann metrics and have therefore identical null geodesics, found as
\beq
\zeta_{+}^{\prime\prime}+\left(1+\frac{1}{4\theta^{2}}\right)\zeta_+ = 0\, \Rarrow
\zeta_+(\theta) =\sqrt{\theta}\Big(\alpha_{1}J_{0}(\theta)+\alpha_{2}Y_{0}\left(\theta\right) \Big)\,
\label{zetathetageo}
\eeq
($\alpha_1$ and $\alpha_2$ are constants), consistently with \eqref{J0Y0eq} and \eqref{importmotion}, respectively. See fig.\ref{zetafigs}i.

The metric $d\Sigma_+^2$ in \eqref{zetathetasigmamet} is in fact the Bargmann metric for the potential
\beq
\Upsilon_{+}(\zeta,\theta) = \frac{1}{2} \zeta_{+}^2 + \frac{1}{8\theta^2}\,\zeta_{+}^2\,,
\label{thetapot+}
\eeq
which describes a unit-frequency(-and-mass) oscillator combined with one with
``inverse-square-in-$\theta$-time''  frequency \cite{Conf4GW,Andr18,AndrPrenc}.

We record for later use with no details that the above procedure works \emph{with the same time redefinition} \eqref{thetat} as in the attractive case, also in the repulsive case with frequency $\omega_-^2 =-e^{-t}$ cf. \eqref{repfreq}, providing us with the analogous formulae
\besub
\begin{align}
&\int\!L_-dt =
\int\!\left\{
\half{\dot{x}}_-^2 - \half{\omega_-^2}x_-^2\right\}dt =
- \half\int\!\theta\left\{\frac{{x'}_-^2}{2\,}+\frac{1}{2}x_-^2\right\}d\theta\,,
\label{thetaact-}
\\[4pt]
&x_- = {\theta}^{-1/2}\,\zeta_-
\aand
\sigma_- = -2s_- - \frac{\zeta_-^2}{4\theta}\,,
\label{zetathetasigma-}
\\[6pt]
&ds_{osc-}^2=\frac{1}{\theta}\,d\Sigma_-^2\,,
\;\;\;
d\Sigma_-^2=d\zeta_-^{2}+2d\theta d\sigma_--\Big(-1+\frac{1}{4\theta^{2}}\Big)\zeta_-^{2}\,d\theta^{2}\,,
\label{zetathetasigmamet-}
\\[6pt]
&{\zeta}_-^{\prime\prime}+\left(-1+\frac{1}{4\theta^{2}}\right)\zeta_-= 0\,
\Rarrow
\zeta_- =\sqrt{\theta}\Big(\beta_{1}I_{0}(\theta)+\beta_{2}K_{0}\left(\theta\right)\Big)\,,
\label{zetathetageo-}
\\[6pt]
&\Upsilon_-(\zeta_-,\theta) = -\frac{1}{2} \zeta_-^2 +
\frac{1}{8\theta^2}\,\zeta_-^2\,.
\label{thetapot-}
\end{align}
\label{repformulas}
\esub
The first terms
in \eqref{thetapot+} and \eqref{thetapot-}  represent attractive/repulsive oscillators with constant frequency-squares $\pm1$,
\beq
\omega^2(t) \big(\cabove(t)\big)^2=1
\quad\text{\small resp.}\quad
 \omega^2(t) \big(\cabove(t)\big)^2=-1\,,
\label{specot}
\eeq
consistently  with \eqref{UpfromV}.
 The second terms are induced in turn by $t\to \theta$ in \eqref{thetat} which works equally for both cases.

The new coordinates  $\big(\zeta_{\pm}, \sigma_{\pm}\big)$ in \eqref{zetathetasigma+} and \eqref{zetathetasigma-} were introduced, separately, by  the rule \eqref{txtauxi}, applied to the redefinition $t \to \theta$ in \eqref{thetat} instead of $t \to \tau$ in \eqref{expofaketime}-\eqref{reptau}~\footnote{
Direct application of sec.\ref{Gibbsec} would run into difficulties~: $\cabove(t)\;= - {2}/{\theta} < 0 $ below the square root to be taken in \eqref{txtauxi}. Arguing however that squaring the quadratic potential would yield a merely negative but  real result, \eqref{thetapot+} and \eqref{thetapot-} would be obtained. }.
 Transforming the potential to these  $\Upsilon_{\pm}\neq0$ forms is, in both cases, advantageous for the good reason that they admit analytical solutions, \eqref{zetathetageo} and \eqref{zetathetageo-}, respectively.

\begin{figure}
\hskip1mm
\includegraphics[scale=0.34]{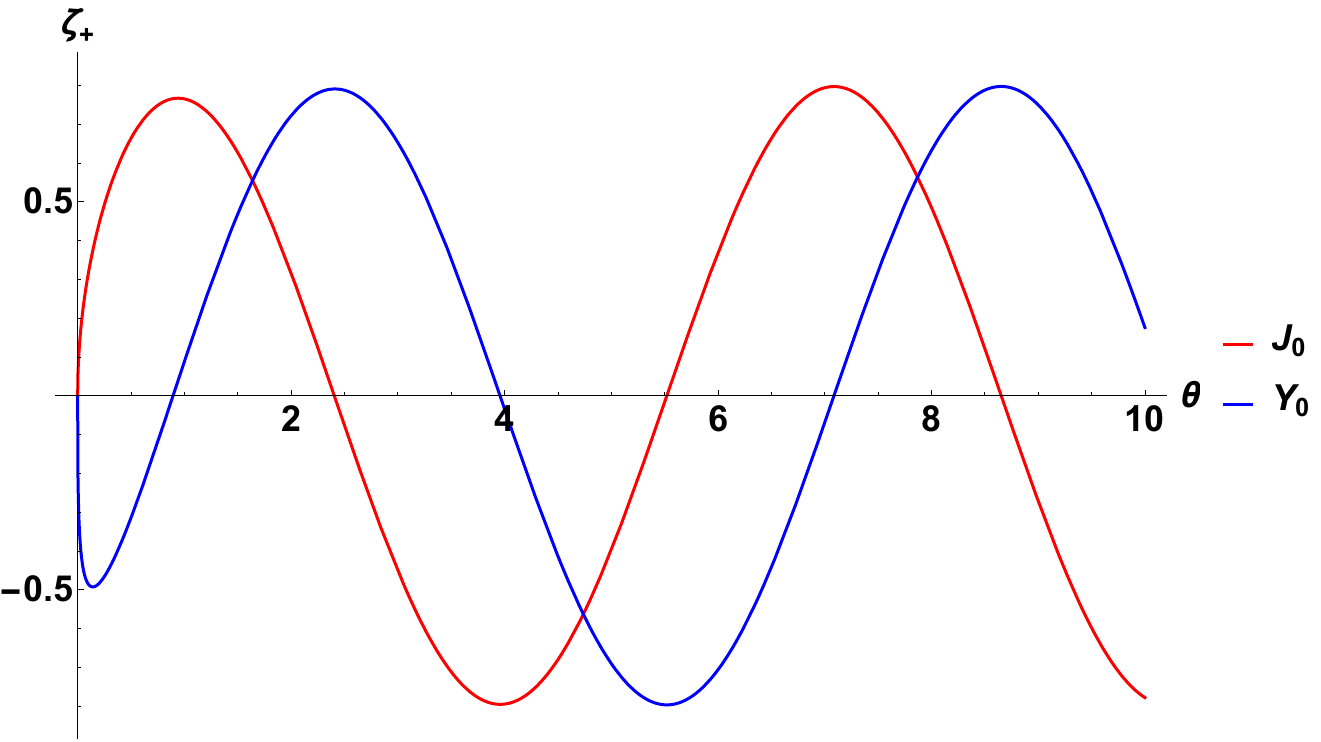}\qquad
\includegraphics[scale=0.33]{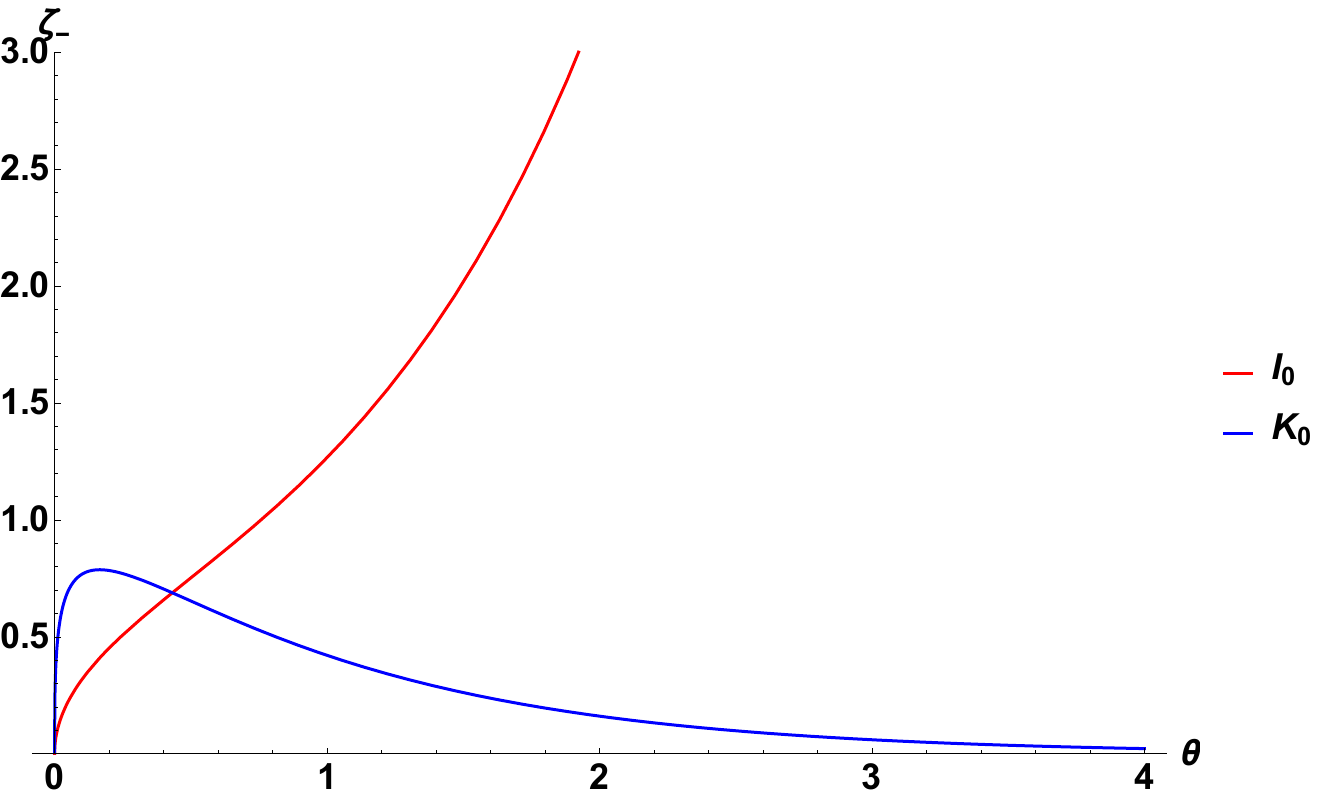}
\\
\vskip-2mm\hskip-20mm
(i) \hskip70mm (ii)\\
\vskip-4mm
\caption{\small \textit{Motions in redefined coordinates $(\zeta_{\pm}, \theta)$ for (i)
the attractive \eqref{thetat} and resp. (ii)  the repulsive \eqref{repfreq} frequencies. For large values of $\theta$ the perturbation term falls off and we are left approximately with harmonic motion~: $\zeta_+(\theta)$ is approximately a shifted sinus/cosinus, and $\zeta_-(\theta)$ is approximately an exponential. (Caveat:  the scales of (i) and (ii) are different~.)
}
\label{zetafigs}
}
\end{figure}

\section{Applications to gravitational waves}\label{GWsec}

Attention has been devoted so far mostly to cases which could be mapped to a free particle \cite{Niederer73,Takagi,BurdetOsci,Andr18}. In particular, when the Niederer map \eqref{NiederTX} works. These cases are distinguished by their conformally flat Bargmann descriptions \cite{Eisenhart, DBKP,DGH91,DHP2,Silagadze}, which
implies isotropy. However plane gravitational waves are vacuum Einstein solutions which correspond to \emph{anisotropic} oscillators whose Bargmann spaces are \emph{not} conformally flat, and therefore the original Niederer map \eqref{NiederTX} does not work. Luckily, the time redefinition technique outlined in sec.\ref{Gibbsec} is not limited to isotropic oscillators and to conformally flat Bargmann spaces, as we illustrate it below  by physical examples.

\subsection{Linearly polarized \GW with ``time''-dependent profile}\label{linpolSec}

Let us first consider the linearly polarized plane \GW  described by the $d+2=4$ dimensional Brinkmann metric \eqref{genBmetric},
\beq
d{\bx}^{2}+2dudv-\omega^{2}(u)\left(x_+^{2}-x_-^{2}\right)du^{2}\,,
\label{Udepo}
\eeq
where $\bx = \binom{x_+}{x_-}$ is the transverse coordinate and we changed our notation, $t \leadsto u$, (a light-cone coordinate), as it is usual in the \GW context. The relative minus between the coordinate-squares is dictated by the vacuum Einstein equations which are then satisfied for arbitrary $\omega^2(u)$.
This metric is \emph{not conformally flat} because of the relative minus.

From the Bargmann point of view \cite{DBKP,DGH91} $u$ is non-relativistic time and the metric \eqref{Udepo} is obtained by putting together two uncoupled
oscillators, one of which is attractive and the other is repulsive. It is therefore \emph{anisotropic} \cite{ZHAGK}.
The simplest example is obviously
$\omega=\omega_0=\const$, when the metric reduces to that of Brdi\v{c}ka \cite{Brdicka,Conf4GW}.

\goodbreak
To get more insight we consider the toy model given by \eqref{Udepo} with the ``time''-dependent frequency we considered in sec.\ref{ExpOex},
\beq
\omega(u)=e^{-u/2}\,,
\label{expfreq}
\eeq
 --- but now in $d=2$ transverse dimensions and with the relative minus dictated by the vacuum Einstein equations duly taken into account. When $u\to 0+$ the metric is essentially that of Brdi\v{c}ka \cite{Brdicka}, and becomes in turn Minkowskian when $u \to \infty$.
The transverse components of the geodesic equations,
\beq
\Big(\begin{array}{c}
\ddot{x}_+ \\ \ddot{x}_-
\end{array}\Big)
 = \Big[K_{ij}(u)\Big]
\Big(\begin{array}{cc}
x_+ \\ x_-
\end{array}\Big),
\qquad
\Big[K_{ij}(u)\Big] = -e^{- u}
\Big[\begin{array}{rr}
 1 &0
 \\
0 & - 1
\end{array}\Big]
\label{2dexpeqmot}
\eeq
(with $\dot{\{\,\cdot\,\}}= d/du$)
are separated~: in \eqref{2dexpeqmot}
 we recognize those \emph{two uncoupled S-L equations} with \emph{opposite frequency-squares},
we solved in sec.\ref{ExpOex}. By  \eqref{J0Y0eq} and \eqref{I0K0eq}
 the solution is therefore (with $t \leadsto u$),
\beq
\left(\begin{array}{cc}
x_{+}(u) \\[4pt]
x_{-}(u)
\end{array}\right)
=
\left(\begin{array}{ccc}
\alpha_{1}J_{0}\left(2e^{-u/2}\right)
&+&\alpha_{2}Y_{0}\left(2e^{-u/2}\right)
\\[4pt]
\beta_{1}I_{0}\left(2e^{-u/2}\right)
&+&
\beta_{2}K_{0}\left(2e^{-u/2}\right)
\end{array}\right)\,.
\label{doublesol}
\eeq

So far so good.
But how does this fit into our framework ? The first step would be to consider a conformal time redefinition $t\to\tau$. However while \eqref{txtauxi} works for an \emph{isotropic} oscillator, we get into trouble in the \emph{anisotropic} case we are studying here~:
as emphasised in sec.\ref{ExpOex},
 eqns. \eqref{expofaketime} and
\eqref{reptau} yield \emph{two different} fake times,
one for each component.  Thus the approach of sec.\ref{Gibbsec} does not apply in its conformally-flat version~: choosing either $t \to \tau_+$ or $t \to \tau_-$ would flatten one, but not the other sector.

Can we do better ? Now we show that our \emph{alternative time redefinition}
\beq
u \to \theta(u)=2e^{-u/2}
\label{thetatbis}
\eeq
smuggled in as \eqref{thetat} \emph{for  both components does save} the situation.
\beq
V(x,u)=\left\{\barraynb{rlll}
\half\omega^{2}(u)x^2&\text{\small attractive}
\\[9pt]
-\half\omega^{2}(u)x^2&\text{\small repulsive}
\earraynb\right.
\label{omegatheta}
\;\where\;
\omega^{2}(u)=\frac{\theta^{2}(u)}{4}=(\cabove(u))^{-2}\,
\eeq
which is \eqref{specot}.
Then setting
\beq
\label{zeta+-}
\zeta_{\pm}=\sqrt{\theta}\,x_{\pm}
\aand
\sigma = \Big(-2v-\frac{\bzeta^2}{2\theta}\Big)
\where
\bzeta=\binom{\zeta_+}{\zeta_-}
\,, %
\eeq%
the metric \eqref{Udepo}--\eqref{expfreq} is rewritten  as,
\besub
\begin{align}
ds_{LP}^{2}&=\frac{1}{\theta}\,d\Sigma^2,
\;\where
\\[2pt]
d\Sigma^2&=d\bzeta^2+2d\theta d\sigma
-2\left\{\half\left(\zeta_+^{2}-\zeta_-^{2}\right)+\frac{\bzeta^2}{8\theta^{2}}\right\}
d\theta^{2}\,.
\end{align}
\label{LPmetric}
\esub
Apart of its anisotropy, the  metric  \eqref{LPmetric} follows the pattern we had found in sec \ref{thetasec}~: the effective potential is the sum of a
($\theta$-)time-independent but  \emph{anisotropic} Brdi\v{c}ka term \cite{Brdicka,Conf4GW}, combined (``perturbed'')
by a second, \emph{isotropic} oscillator with a $\theta$-time dependent frequency, induced by the Schwarzian cf. \eqref{UpfromV},
\beq
\smallover{1}/{4}\calS_\theta(u)=\frac{1}{8\theta^2}\,.
\label{Sthetat}
\eeq

The motion in the transverse plane as a function of $\theta >0$, consistent with \eqref{doublesol}, is shown in fig.\ref{3d-YKfig}.
\begin{figure}
\hskip-5mm
\includegraphics[scale=0.275]{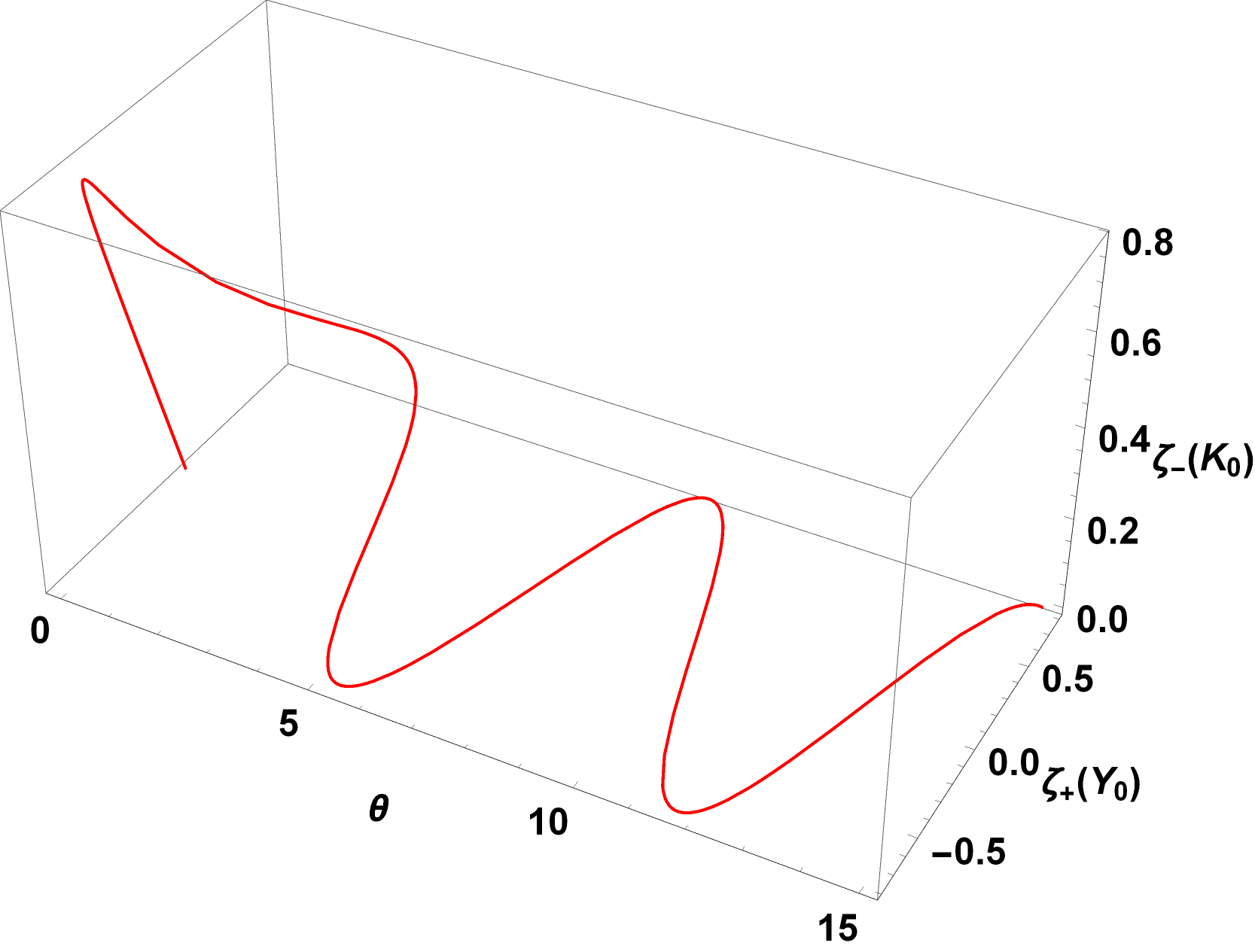}\;
\includegraphics[scale=0.275]{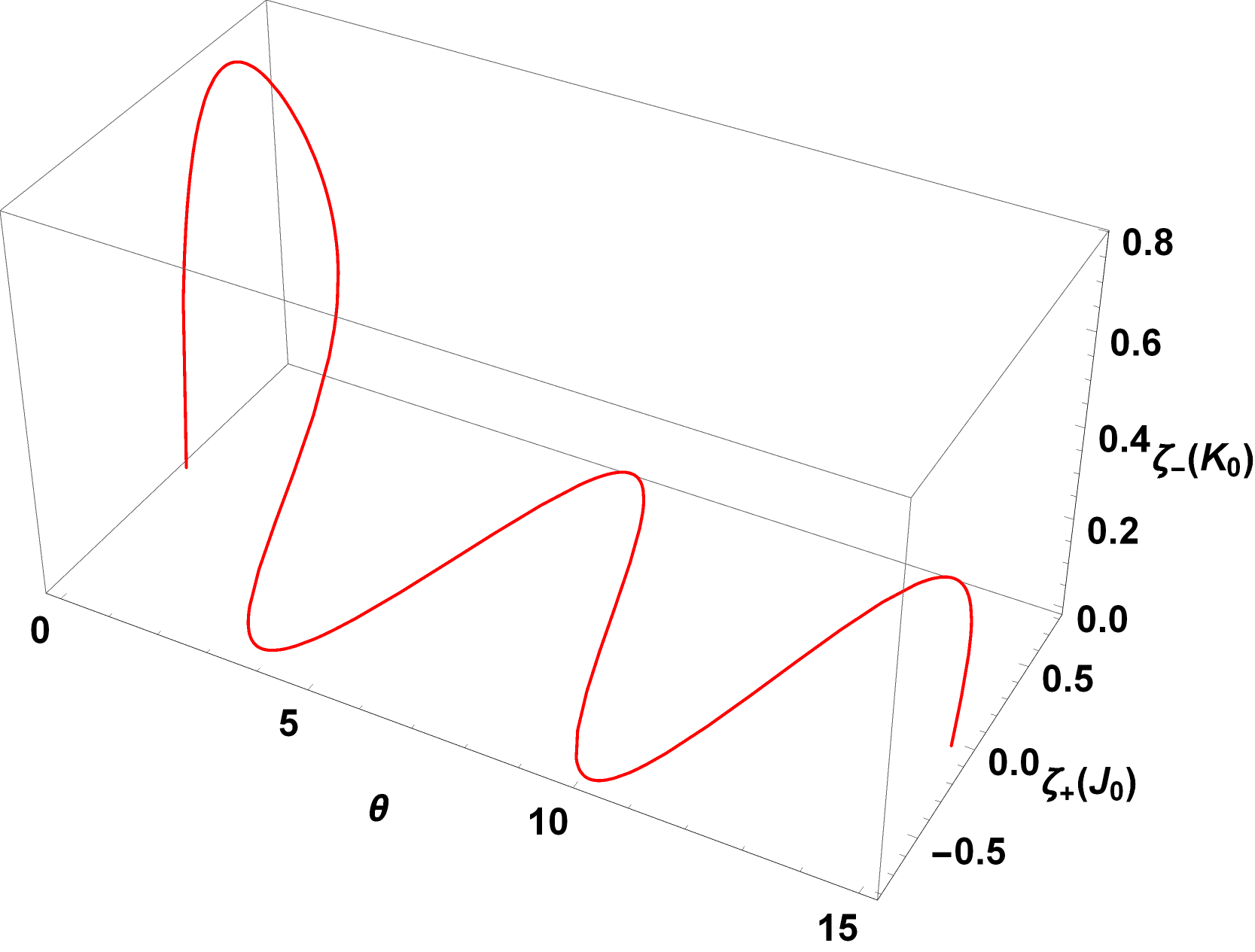}\\
(i) \hskip73mm (ii)\\
\vskip-5mm
\caption{\small \textit{
 Motion in the transverse plane for the choices (i) $\zeta_{+} \sim Y_0$ and for  (ii) $\zeta_{+} \sim J_0$, completed, in both cases, with  $\zeta_- \sim K_0$. For large $\theta$ $\zeta_-(\theta)$ falls of exponentially and the $\zeta_+$ component moves approximately along a shifted sinus curve. Two more figures could be produced by other pairings of the curves taken from fig.\ref{zetafigs}.}
\label{3d-YKfig}
}
\end{figure}

In conclusion, the time redefinition $u\to\theta$  takes the \GW \eqref{Udepo}
conformally into the perturbed Brdi\v{c}ka metric. The conformal factor is ${\theta^{-1/2}}$.

A look at \eqref{UpfromV} applied to both (uncoupled) components in \eqref{Udepo} tells us that in general a constant-coefficient first term arises when
$ (\cabove(t))^2\omega^2(t)=C$
for some real constant $C$, cf. \eqref{specot}.

\goodbreak
\subsection{Circularly polarized plane gravitation waves}\label{CPGW}

 A physically relevant illustration is given by \emph{Circularly Polarized Gravitation Waves} \cite{exactsol}.
Let us consider a Lukash gravitational wave proposed  to study the anisotropy of the Universe \cite{LukashJETP75,exactsol}, whose metric is in Brinkmann coordinates, \vspace{-2mm}
 \besub
 \begin{align}
ds^2_{L}&=d\bx^2 + 2 du dv
+ K_{ij}(u)x^ix^j,
\\[6pt]
K_{ij}(u)x^ix^j&=- \frac{C}{u^{2}}\left\{
\cos\Big(2\kappa\ln(u)\Big)
\,
\big[(x^1)^2-(x^{2})^2\big]
\,-\,
2\sin\Big(2\kappa\ln(u)\Big)
 \,x^1x^2\right\}\,du^2\,,
\end{align}
\label{BLukash}
\esub
where $\bx=(x^1,x^2)$ are transverse coordinates and $u>0$ and $v$ are light-cone coordinates. $C$ is the amplitude and $\kappa=\const$ is the frequency of the wave \cite{LukashII}.
Finding its geodesics is reduced to solving the 2-dimensional Sturm-Liouville system for the transverse coordinates \cite{Elbistan:2020ffe},
\beq
\left(\begin{array}{cc}
\ddot{x}^1 \\[3pt] \ddot{x}^2
\end{array}\right)
 = -\frac{C}{u^2}
 \left(\begin{array}{rr}
 \cos (2\kappa \ln u) &- \sin (2 \kappa \ln u)
 \\[3pt]
- \sin (2 \kappa\ln u) & - \cos (2\kappa \ln u)
\end{array}\right)
 \left(\begin{array}{cc}
x^1 \\[3pt] x^2
\end{array}
\right)\,
\label{LukashBXY}
\eeq
where now $\dot{\{\,\cdot\,\}}= d/du$.

From the ``Bargmannian'' point of view \cite{Eisenhart,DBKP,DGH91} $u$ is non-relativistic time; the metric \eqref{BLukash} and the equations \eqref{LukashBXY}, respectively, describe the motion of an anisotropic oscillator with coupled  components in the plane with rather complicated ``time''-dependent frequencies.
However switching  to  new  coordinates $(u,\bx) \to (\theta, \bxi)$ by following the rule  \eqref{txtauxi} with the choice,
\beq
u = e^{\theta}\;\Leftrightarrow\; \theta = \ln u\,,
\quad
\bx=e^{\theta/2}\bxi
\where
\bxi={\tiny \barray{c}\xi\\ \eta\earray}\,,
\label{lnchange}
\eeq
eqn. \eqref{LukashBXY} becomes,
\beq\bigbox{
\frac{\;\; d^2}{d\theta^2}
\barray{c}\xi\\\eta\earray=
\underbrace{- C
\barray{rr}
 \cos(2\kappa \theta) &- \sin (2\kappa \theta)
 \\
- \sin(2\kappa \theta) & - \cos (2\kappa \theta)
\earray
\barray{c}\xi\\[3pt] \eta\earray}_{\text{anisotropic  CPP}}
+
\underbrace{\frac{1}{4}\barray{c}\xi\\ \eta\earray}_{\text{isotropic inverted osc}}\,,
}
\label{lnk}
\eeq
which describes the motion in the more familiar
 \emph{Circularly Polarized Periodic} (CPP) wave with constant frequency $\omega= 2\kappa$ \cite{exactsol,POLPER,KosinskiCPol}, combined (``perturbed'') by a repulsive isotropic planar oscillator \cite{LukashII}.

Eqn.  \eqref{lnk}  follows the  pattern found  in secs.\ref{thetasec} and \ref{linpolSec} and
 fits into the framework outlined in sec.\ref{Gibbsec}~: the CPP and linear terms in \eqref{lnk} correspond (by $S_{\theta}(u)=-\half$) to those in \eqref{Upsipot}.

The system \eqref{lnk} can  be solved analytically. The  coordinate transformation at constant $\theta$ \cite{POLPER,IonTrap,BB}
\begin{equation}
\bxi=
R_{\kappa\theta}\,\bzeta
\where
\bxi=\binom{\xi}{\eta},\quad
R_{\kappa\theta}= {\small
\Big(\begin{array}{cc}
\;\;\cos \kappa \theta
 &  \sin \kappa \theta
 \\
 - \sin \kappa \theta
& \;\cos\kappa \theta
\end{array}\Big)}\,,%
\quad
\bzeta =\binom{\alpha}{\beta}
\label{Lukrotframe}
\end{equation}%
carries \eqref{lnk} to a ``Coriolis'' form with \emph{constant coefficients},
\beq
\medbox{\barraynb{l}
\ccabove(\alpha)
+2\kappa \cabove(\beta)
-\Omega_-^2 \,\alpha =0\,
\\[4pt]
\ccabove(\beta)
-2\kappa\, \cabove(\alpha)
-\Omega_+^2 \, \beta =0\,
\earraynb
\label{abeqmot}
}\eeq
where  $\cabove(\{\,\cdot\,\})$ means $d/d\theta$ and
\begin{eqnarray}
\Omega_{-}^2=\smallover{1}/{4}+\kappa^2-C, \aand
\Omega_{+}^2=\smallover{1}/{4}+\kappa^2+C\,.
\label{Omega-+}
\end{eqnarray}
The new frequencies $\Omega_{\pm}$ which combine the strength and frequency of the Lukash wave make the anisotropy manifest. The equations can nevertheless be separated, and then integrated  by using a powerful technique referred to chiral decomposition \cite{Alvarezys,Alvarez07fw,IonTrap,BB,LukashII}.

The metrics behave similarly~:  E-D lift of \eqref{lnchange} completed with
$v = \nu - \smallover{1}/{4}\bxi^{2}$
 is a conformal transformation between the Lukash  and  (perturbed) CPP spacetimes, as seen by presenting $ds_L^{2}$ as,
\besub
\begin{align}
ds_L^{2}&= e^{\theta}d\Sigma^2\, \with
\label{LconfCPP}
\\
d\Sigma^2&
=d\bxi^{2}+2d\theta d\nu-2\underbrace{\left\{
\overbrace{-C\Big(\half(\xi^{2}-\eta^{2}) \cos2\kappa \theta-\xi\eta
\sin 2\kappa \theta \Big)}^{CPP}\quad
\overbrace{\,-\;\smallover{1}/{8}\,\bxi^{2}}^{osc}\, \right\}}_{\Upsilon(\bxi,\theta)}\,d\theta^{2}\,.
\label{perCPP}
\end{align}
\label{Lukrescaled}
\esub
In the braces we recognize the redefined potential $\Upsilon(\bxi,\theta)$, decomposed as in \eqref{Upsipot}.
Conformal transformations permute the null geodesics and  therefore the latter are identical for the Lukash and the rescaled metrics, $ds_L^2$ and $d\Sigma^2$, respectively. The coefficient of the 2nd term in \eqref{perCPP} is consistent with the Schwarzian in \eqref{Upsipot}.
Accordingly,
\beq
\Big\{\half \,\dot{\bx}^2-K_{ij}(u)x^i x^j\Big\}\,du
=
e^{\theta}\,
\Big\{\half \,
{\cabove(\bxi)}^2-\Upsilon(\bxi,\theta)\Big\}\,d\theta\,.
\label{UpsiKpot}
\eeq

The metric \eqref{Lukrescaled} could also be brought to yet another form by switching to the rotating coordinates $\bzeta=\binom{\alpha}{\beta}$ introduced in \eqref{Lukrotframe},
\beq
d\Sigma^2=d\bzeta^2+2d\theta \left[d(\nu
+\kappa\left( -\alpha\cdot d\beta+\beta\cdot d\alpha\right)\right]
+2\Big\{\half\Omega_-^2\alpha^2+\half\Omega_+^2\beta^2
\Big\} d\theta^2\,,
\label{Lrotcoords}
\eeq
which is the Bargmann metric of
 a charged   planar oscillator
in combined but ($\theta$-)\emph{``time''-independent} vector and anisotropic scalar potentials
cf. \eqref{Bvecpot},
\beq\barraynb{clcll}
\bA(\bzeta)\cdot d\bzeta &=&
\kappa\left( -\alpha\cdot d\beta+\beta\cdot d\alpha\right)&
&\text{\small vector potential}
\\[4pt]
V(\bzeta)&=& \half\Omega_-^2\alpha^2+ \half\Omega_+^2\beta^2\
&\qquad
&\text{\small scalar potential}
\earraynb
\label{simplestdSigma}
\eeq
The variational equations are \eqref{abeqmot}. Following the redefinition \eqref{lnchange} by the rotation \eqref{Lukrotframe} adapts \eqref{txtauxi} to our present, anisotropic case,
\beq\medbox{
t = t(\theta),\qquad
\bx=e^{\theta/2}R_{\kappa\theta}\,\bzeta\,.}
\label{Lukpos}
\eeq

\section{Epilog~: hint at quantum aspects}\label{QM}

The classical aspects studied in this note could readily be extended to Quantum Mechanics. For quadratic systems the semiclassical expression of the propagator is exact, and can be evaluated using classical data  only \cite{FeynmanHibbs,Schulman,KLBbook}. Many alternative approaches  exist, e.g. \cite{Curtright}, which uses quantum operator identities.

For an harmonic oscillator with  constant frequency $\omega_0$ a lengthy calculation yields the exact propagator \cite{FeynmanHibbs,Schulman,KLBbook,BurdetOsci,HFeynman,ZZH}
\besub
\begin{align}
&K_{osc}(x'',t''|x',t')=
\left[\frac{\omega_0}
{2\pi i\hbar\,\sin\big|\omega_0(t''-t')\big|}\right]^{\frac{1}{2}} \times \,e^{-i\frac{\pi}{4}(1+2\ell)}\times
\\[6pt]
&\qquad\exp\left\{\frac{i{\omega_0}}{2\hbar\,\sin\big[{\omega_0}(t''-t')\big]}\left[
\big({x''}^2+{x'}^2\big)\cos\big[{\omega_0}(t''-t')\big]-2x''x'
\right]\right\},
\end{align}
\label{osciprop}
\esub
which includes also the {Maslov correction}  \cite{Maslov} highlighted by $\ell = {\rm Ent} \big[\frac{\omega_0(t''-t')}{\pi}\big]\,,
$
where ${\rm Ent}[\,\cdot\,]$ denotes the integer part \cite{HFeynman,BurdetOsci}. This formula can also be derived \cite{HFeynman} by using the subtleties of the Niederer map (whose inverse is multivalued) \cite{BurdetOsci,ZZH}.

\goodbreak
\section{Conclusion and outlook}

  Gravitational waves are at the center of current interest. In suitable coordinates they may correspond to anisotropic oscillators with time-dependent frequency.

  The classical Niederer transformation  \cite{Niederer73} and its subsequent generalizations \cite{Takagi,BurdetOsci,JI85,DBKP,DGH91,DHP2,PerrinBD,GWG_Schwarz,Andr18,Andr2014,ZHAGK,AndrPrenc,Galajinsky,Inzunza,ZZH,Masterov,Silagadze, Curtright,Arnold,Aldaya,CDGH}
 relate \emph{isotropic} oscillators to free particles and
 apply therefore only to spacetimes which are conformally flat --  excluding \GWs. Generalizing the conformal redefinition proposed by Gibbons \cite{GWG_Schwarz} led us to extending the correspondence to \emph{anisotropy} so that it  applies to plane gravitational waves.

In detail, in secs.\ref{ExpOex} and \ref{linpolSec} we argue using a toy model that while the usual time redefinition (we refer here to as ``$\tau$-type'') fails to work
for our anisotropic system, a less demanding ``$t \to \theta$ type'' redefinition, our  \eqref{thetat}, maps it --- not to a free, but to one which we happen to be able to solve.

In the physically relevant
 Lukash \cite{LukashJETP75,LukashI,LukashII} versus CPP  \cite{Ehlers,exactsol} context Gibbons' formula \eqref{txtauxi} replaced  by \eqref{Lukpos} relates two \emph{anisotropic and time-dependent} \GWs.

 Similar questions have been discussed also in the Newton-Cartan framework in   \cite{DuvalThese,DuvalCosmo} we do not consider here.

\vskip-5mm
\kikezd{Outlook for further research}.
The Niederer-Takagi-Gibbons approach followed in our paper was parallelled recently and independently by a series of interesting developments which would deserve further research.

Firstly, the planar anisotropic harmonic oscillator (and its $d>2$ generalisations) with explicit rotational symmetry as a particle model with non-commutative coordinates was discussed in \cite{Alvarezys}. The exotic planar oscillator with the Hamiltonian and angular momentum interchanged also appeared in \cite{Alvarez07fw}.

The Landau problem appears as a particular case which separates the Euclidean and Minkowskian phases in the conformal generation of an exotic rotationally invariant harmonic oscillator \cite{Inzunza}. Working in the plane with one of the two frequencies putting to zero corresponds to the Landau problem, which could probably be  included into the schemes discussed here.

In refs.  \cite{Inzunza20} and in  \cite{Inzunza21} a generalisation of the conformal bridge transformation to cosmic string backgrounds was considered. One can wonder if it is possible to extend our results here to gravitational waves in conical space-times.


This paper grew out from the talk given by one of us (PH) on 20 Dec. 2021 at the Miami 2021 Winter Conference, see
 https://cgc.physics.miami.edu/Miami2021.html.

\vspace{-5mm}
\begin{acknowledgments}\vskip-4mm
 We are indebted to Gary Gibbons for advice, to Mahmut Elbistan for discussions and to Ivan Masterov for correspondence.
This work was partially supported by  the National Natural Science Foundation of China (Grant No. 11975320).
\end{acknowledgments}
\goodbreak

\end{document}